\documentclass[10pt,conference]{IEEEtran}
\usepackage[left=0.70in, right=0.73in, top=0.7in, bottom = 0.98in]{geometry}
\usepackage{graphicx}
\usepackage{setspace}
 \usepackage{comment}
\usepackage{adjustbox} 
\usepackage{enumitem}
\usepackage[noadjust]{cite}
\usepackage[linesnumbered,ruled]{algorithm2e}
\usepackage{amsmath}
\usepackage{lipsum}
\usepackage{mathtools}
\usepackage{cuted}
\usepackage{array}
\usepackage{subfigure}
\usepackage{amssymb}
\usepackage{caption}
\usepackage{multirow}
\usepackage{makecell}
\usepackage{algorithmic}
\usepackage{xcolor}
\def\BibTeX{{\rm B\kern-.05em{\sc i\kern-.025em b}\kern-.08em
ThedeltaT\kern-.1667em\lower.7ex\hbox{E}\kern-.125emX}}

\SetCommentSty{mycommfont}
\usepackage{array,multirow,makecell}
\SetKwInput{KwInput}{Input}    
\SetKwInput{KwRepeat}{Repeat} 
\SetKwInput{KwOutput}{Output}    

\begin{document}


\title{Joint Beamforming and Antenna Placement Optimization in Pinching Antenna Systems with user Mobility: A Deep Reinforcement Learning Approach}
\author{Ali Amhaz, Mohamed Elhattab, Chadi Assi and Sanaa Sharafeddine \vspace{-0.6cm}}

\maketitle

\begin{abstract} 
Recently, the pinching antenna systems (PASS) have attracted significant attention due to their ability to exploit dynamically reconfigurable pinching points along waveguides for flexible signal transmission. However, existing work largely overlooks user mobility although the optimal pinching configuration is highly dependent on the user’s location and must be continuously adjusted. In this work, we investigate a PASS-enabled system model in which a base station (BS) serves a mobile user. We formulate an optimization problem that aims to maximize the user’s average sum rate over a predefined time horizon while satisfying quality-of-service (QoS) constraint. This objective is achieved by jointly optimizing the beamforming vector at the BS and the pinching locations along the waveguides. Nevertheless, the resulting problem is highly non-convex and challenging to solve using conventional optimization techniques due to the intricate coupling among variables. The difficulty is further exacerbated by environmental randomness arising from user mobility and a probabilistic blockage model. This reveals a key engineering challenge: the performance gains of PASS critically rely on the ability to track or predict user trajectories in real time. To address these challenges, we adopt a deep deterministic policy gradient (DDPG) approach within a reinforcement learning framework, which is well-suited for continuous state and action spaces. Finally, extensive simulations are conducted to validate the proposed approach and demonstrate the importance of real-time configurability.

\end{abstract}

\section{Introduction}
 The sixth-generation (6G) of wireless networks is anticipated to trigger a revolution of technological advancement far beyond what earlier generations have attained. Fueled primarily by breakthroughs in artificial intelligence (AI) and automation, the accelerated growth of emerging services and applications will place stricter requirements on future systems, including significantly higher data throughput,
 enhanced spectral and energy efficiency, minimal latency, and better reliability \cite{9349624}. 

From the second to the fifth generation of wireless networks, multiple-input multiple-output (MIMO) systems have served as a fundamental pillar in enhancing performance by increasing the available degrees of freedom (DoFs) \cite{9733081}. By employing multiple antennas at both the transmitter and receiver, MIMO significantly improves spectral efficiency, boosts data rates, and enhances signal reliability. These gains are achieved by exploiting spatial diversity and multiplexing, which help mitigate interference and extend coverage.  MIMO has progressed into more advanced variants, including massive MIMO, gigantic MIMO, and, more recently, continuous aperture arrays (CAPA) \cite{11222687}. While these approaches can further enhance spectral efficiency, they also introduce greater costs, such as increased computational complexity, substantial channel estimation overhead, and higher implementation expenses.

In the last couple of years, pinching antennas systems (PASS) have gained considerable attention owing to their capability to dynamically alter the wireless transmission environment following an analogous direction pursued in earlier technologies such as reconfigurable intelligent surfaces and movable antennas. Unleashed by NTT DOCOMO at the Mobile World Congress (MWC) 2021, PASS demonstrates a clever reconfigurable antenna system established over dielectric waveguide principles in which electromagnetic waves traverse low-loss dielectric waveguides, while what are referred to as “pinching points” allow controlled radiation of signals into free space at specific locations along the guide.

The research community focused on optimizing the placement of these pinching points, demonstrating performance gains compared to traditional fixed-position antenna systems in terms of enhanced user data rates \cite{11222687}. Given their strong potential in manipulating wireless channels, especially large-scale fading over extended distances, PASS would provide better or complimentary performance to conventional MIMO systems that often suffer from reduced effectiveness due accompanied challenges. In \cite{10896748}, the authors exploited PASS technology to maximize users’ achievable data rates by identifying optimal pinching point locations and showed that this approach outperforms traditional fixed-antenna benchmarks.

Although PASS proved to acquire superior performance over traditional antenna systems, most existing studies  assume static or quasi-static user locations with idealized propagation conditions limiting their applicability in realistic mobile environments \cite{11364174}. In practical scenarios, users are predicted to be highly mobile, with movement patterns often modeled using stochastic processes such as the random waypoint (RWP) mobility model. Under such scenarios, the channels are rapidly time-varying, due to the users continuous change of direction and speed. This mobility effect is further intensified by blockages in which obstacles such as buildings, vehicles, or even human bodies obstruct the line-of-sight (LoS) and introduce sudden transitions between LoS and NLoS states. The researchers in \cite{6932503} study the effect of blockage on the performance of millimeter-wave (mmWave) systems by relying on a stochastic framework that distinguishes between LoS and NLoS channels, where the results prove that blockage plays a dominant role in shaping network performance.

The combination of these effects together yields non-stationary channels, making it difficult to determine the optimal pinching point configurations over time, degrading the overall performance. As a result, it is essential that the system quickly adapt the pinching placement and beamforming vectors. In light of this and to the best of our knowledge, no work has studied the user mobility scenario in PASS system with blockage model. Hence, we formulate an optimization problem with the aim of maximizng the average sum rate at the end of the time horizon while guaranteeing the quality of service (QoS) constraint for the user. The resulting problem is highly non-convex due to the strong coupling between pinching point locations, beamforming design, and the time-varying channel conditions imposed by mobility and blockage. Moreover, the stochastic nature of the environment and the insufficient prior knowledge about future channel states result in the inapplicability of traditional optimization techniques. To tackle these challenges, we employ a Deep Deterministic Policy Gradient (DDPG) method within a reinforcement learning framework  capable of capturing the system dynamics and adapting decisions in real time and particularly well-suited for handling continuous state and action spaces \cite{10570841}. Finally, extensive simulations are carried out to validate the proposed approach and demonstrate its effectiveness.

\section{System Model}
\begin{figure} 
    \centering    \includegraphics[width=0.45\textwidth]{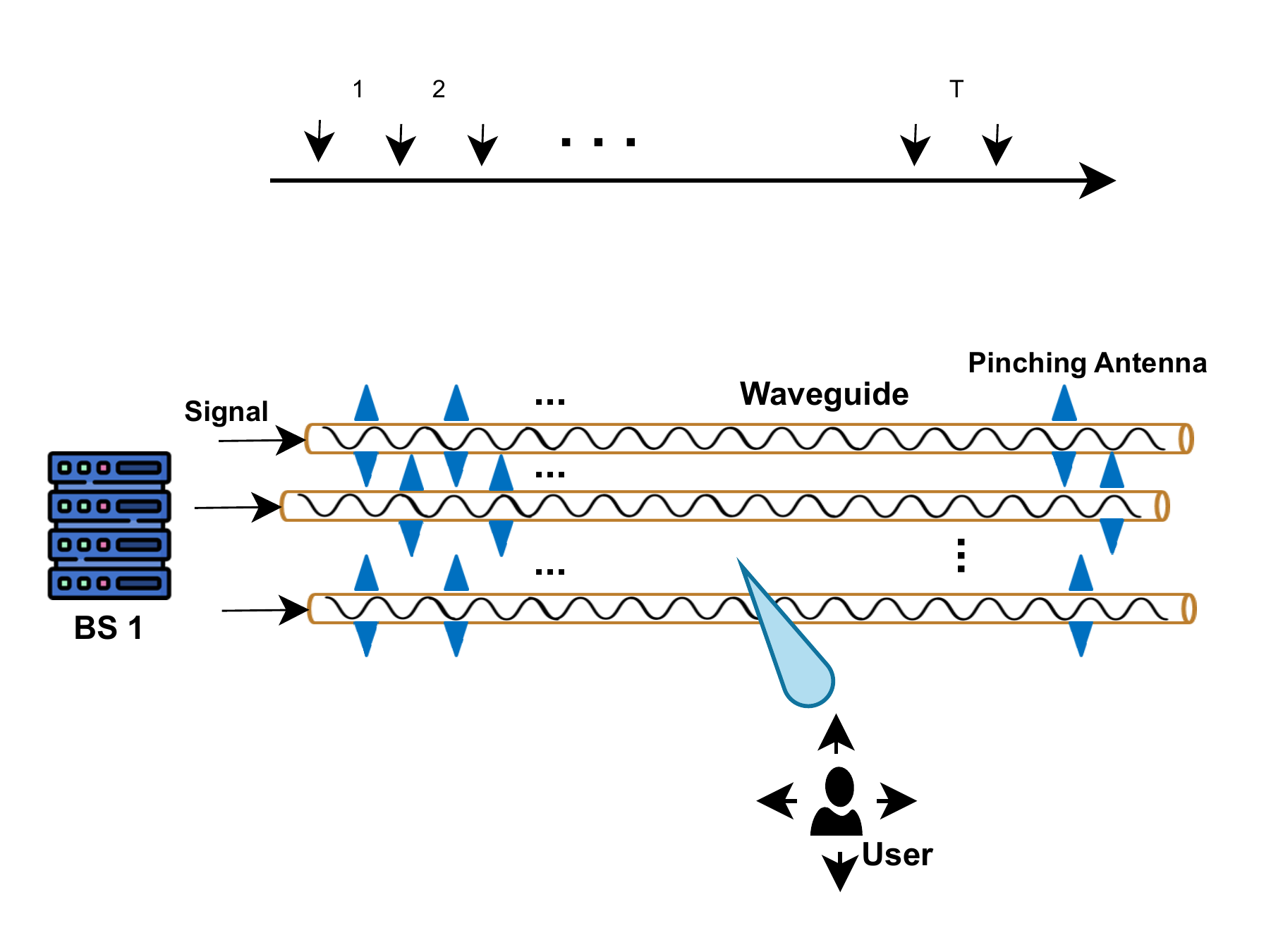}
    \caption{System Model}
    \label{fig:placeholder}
\end{figure}
This work considers a PASS-enabled system in which a base station (BS) serves a mobile user denoted by $k$ with coordinates $\boldsymbol{u}_k^t = [x_k^t, y_k^t, 0]^T$ at the $t$-th time slot, as depicted in Fig. 1. The BS comprises a set of $N$ waveguides defined as $\mathcal{N} = \{1, 2, \dots, n, \dots, N\}$ and connected via RF chains. These waveguides are assumed to be parallel and oriented along the x-axis, with the height of the $n$-th waveguide denoted by $A_n$. For each waveguide $n$, the y-axis coordinate is defined as $y_{n}$, yielding a waveguide that spans from the feeding point $\boldsymbol{O}_{n} = [0, y_{n}, A_{n}]^T$ to the end point $\boldsymbol{f}_{n} = [\mathcal{D}_{n}, y_{n}, A_{n}]^T$ with $\mathcal{D}_{n}$ representing the length of the waveguide. In addition, the $n$-th waveguide acquires a set of $P_{n}$ pinching antennas (PAs) that could be placed in a continuous manner throughout the time slots so that the coordinate of the $p$-th antenna in time slot $t$ can be defined as $\boldsymbol{p}_{n,p}^t = [x_{n, p}^t, y_{n}^t, A_{n}]^T$.


Now, the complex information symbol that correspond to the movable user $k$ can be expressed as $c_k$ satisfying $\mathop{\mathbb{E}} \{\boldsymbol{c}_k\boldsymbol{c}_k^H\} = \boldsymbol{I}_K$. Thus, the signal at the BS can be expressed as follows 
\begin{equation}
    \textbf{s}=\textbf{w}_{k}{c}_{k},
\end{equation}
where $\textbf{w}_{k} \in \mathbb{C}^{N \times 1}$ denotes the beamforming vector. In this work, we assume that the in-waveguide propagation is lossless, yielding the following representation for channel inside the $n$-th waveguide:
%
\begin{align}
&\textbf{g}_{n}=\sqrt{
\delta_{\text{eq}}
}\left[e^{\frac{-2\pi}{\lambda_g}\|\boldsymbol{p_{{n},1}}-\textbf{O}_{n}\|},\dots,e^{\frac{-2\pi}{\lambda_g}\|\boldsymbol{p_{{n},P_{n}}}-\textbf{O}_{n}\|}\right]^T \nonumber \\
&\in \mathbb{C}^{N \times 1},
\end{align}
where $\lambda_g = \frac{\lambda}{n_{\mathrm{eff}}}$ is the guided wavelength. $n_{\mathrm{eff}}$ is the effective refractive index that characterizes the propagation of the wave within the waveguide and $\lambda$ being the wavelength \cite{zhao2025waveguidedivisionmultipleaccess}. In addition, $0 < \delta_{\text{eq}} \leq \frac{1}{P_{n}}$ supports an equal-power ratio model in which the BS's power budget is evenly distributed over the different radiating pinching locations \cite{wang2025modelingbeamformingoptimizationpinchingantenna}. 
Now, we model the wireless channel between the $n$-th waveguide, equipped with
$P_n$ PAs, and user $k$ using a probabilistic blockage
model. To capture the spatial variability of blockage across different PA
positions, we define a Bernoulli random variable $b_{n,p,k} \in \{0,1\}$
for the link between the $p$-th PA of the $n$-th waveguide and user $k$,
where $b_{n,p,k}=1$ demonstrates a total LoS and $b_{n,p,k}=0$ corresponds to a blocked link. The channel vector is denoted as
\begin{equation}
\mathbf{h}_{n,k}
=
\left[
h_{n,1,k}, \dots, h_{n,P_n,k}
\right]^H,
\end{equation}
where each element is given by
\begin{equation}
h_{n,p,k}
=
b_{n,p,k} \, h_{n,p,k}^{\mathrm{LoS}}
+
(1-b_{n,p,k}) \, h_{n,p,k}^{\mathrm{NLoS}}.
\end{equation}
\noindent The probability of having an LoS link is modeled as a distance-dependent
function
\begin{equation}
\Pr\!\left(b_{n,p,k}=1\right)
=
P_{\mathrm{LoS}}(d_{n,p,k})
=
\exp(-\beta d_{n,p,k}),
\end{equation}
where $\beta$ denotes the blockage density parameter and
$d_{n,p,k} = \|\boldsymbol{p}_{n,p} - \mathbf{u}_k\|$ represents the distance
between the $p$-th PA and user $k$.

\subsubsection{LoS Channel Component}

The LoS component of the channel is given by
\begin{equation}
h_{n,p,k}^{\mathrm{LoS}}
=
\frac{\eta_{\mathrm{mm}}
e^{-j\frac{2\pi}{\lambda}\|\boldsymbol{p}_{n,p}-\mathbf{u}_k\|}}
{\|\boldsymbol{p}_{n,p}-\mathbf{u}_k\|^{\alpha/2}},
\end{equation}
where $\eta_{\mathrm{mm}}$ is the channel gain factor, and $\alpha$ is the path-loss exponent.
\subsubsection{NLoS Channel Component}

When the LoS path is blocked, the channel is modeled as a NLoS component with randomized phases:
\begin{equation}
h_{n,p,k}^{\mathrm{NLoS}}
=
\frac{\tilde{\eta}_{\mathrm{mm}} e^{-j\phi_{n,p,k}}}
{\|\boldsymbol{p}_{n,p}-\mathbf{u}_k\|^{\alpha_{\mathrm{N}}/2}},
\end{equation}
where $\tilde{\eta}_{\mathrm{mm}} \ll \eta_{\mathrm{mm}}$,
$\alpha_{\mathrm{N}} > \alpha$, and
$\phi_{n,p,k} \sim \mathcal{U}[0,2\pi)$ denotes a random scattering phase.

Now, 
we define the signal received at the mobile user as:
%
%
%
\begin{align}
&y_k=\textbf{H}_{ k}^H\textbf{G}\textbf{w}_{k}  + n_k,  
\end{align}
with $n_k$ being the additive white gaussian noise (AWGN) and defining the terms $\textbf{H}_{k}$ and $\textbf{G}$  as follows:
\begin{equation}
    \textbf{H}_{k}=[\textbf{h}_{1,k}^T,\dots,\textbf{h}_{N,k}^T]^T \in \mathbb{C}^{N P_{n} \times 1},
\end{equation}
\begin{equation}
    \textbf{G}=\textrm{diag}[\textbf{g}_1,\dots,\textbf{g}_{N}] \in \mathbb{C}^{N P_{n_b} \times N}.
\end{equation}
Accordingly, the achievable sum rate at the mobile user at the t-$th$ time slot is denoted as follows:
\begin{align}
&R_{k} [t]=\log_2\left(1+\frac{|\textbf{H}_{ k}^H\textbf{G}\textbf{w}_{k}|^2}{\sigma_k^2}\right).
\end{align}
\section{Problem Formulation}
The aim in this work is to maximize the average sum of the achievable rates of the movable user over the different time slots, by jointly determining the transmit beamforming vector, along with the PAs positions ($\boldsymbol{P} = \{ \boldsymbol{p}_{n,p}, \forall n \in \mathcal{N}, 0 < p < P_n\}$), while respecting the QoS requirement. Thus, the problem can be formulated in the following manner:
\allowdisplaybreaks
\begin{subequations}
\label{prob:P1}
\begin{flalign}
 &\mathcal{P}_1: \max_{\substack{\textbf{P}, \,\textbf{w}_k}} \quad \frac{1}{T}\sum_{t=0}^T R_{k}[t], \label{const1}\\
 &\text{s.t.} \quad  (\textbf{w}_k[t]^H\textbf{w}_k[t]) \le P_{BS}, \label{c0}\\
 &\qquad R_k [t]\ge R_{th}, \forall k \in \mathcal{K}, \label{c1}\\
 &\qquad ||\boldsymbol{p}_{n,p}[t] - \boldsymbol{p}_{n,p'}[t]||  \ge D, \; p \neq p', \;  \forall n \in \mathcal{N}, \nonumber \\
 &\qquad  \forall 0 \leq p, p' \leq P_n, \nonumber\\
 &\qquad p_{n} [t]\in \mathcal{D}_{n}. \label{c3}
\end{flalign}
\end{subequations}
where the constraint \eqref{c0} is for the power budget $P_{BS}$ at the BS. \eqref{c1} is to guarantee the QoS requirements for the user $k$. Finally, constraint \eqref{c3} ensures that the spacing between pinching locations exceeds a minimum distance $D$, while also guaranteeing that all pinching points lie within the waveguide.

\section{Solution Approach}
In order to handle the complexity of the formulated problem due to the high coupling between the different variables and to deal with the environmental uncertainty resulting from user mobility model (RWP) along with the blockage model, we utilize a reinforcement learning framework, based on the DDPG algorithm. 

In particular, DDPG is well suited for problems with continuous state and action spaces, enabling the direct optimization of continuous variables such as beamforming vectors and antenna positions without requiring discretization. This makes it especially effective for capturing fine-grained control policies in dynamic wireless environments.

This approach is established relying on an actor--critic architecture, where the actor network, denoted by $\mu$, produces actions based on states, effectively defining the control policy within the environment. Simultaneously, the critic network, defined as $Q$, evaluates the generated policy through a value function. Each component is built exploiting two neural networks: a training network and a target network. Training networks iteratively update their parameters through previously encountered experiences to improve both policy and value estimation, while target networks, expressed as $\mu'$ and $Q'$, provide  updated reference values that maintain training stability. Accordingly, the loss function is defined as follows:
\begin{align}
& l(\theta) = \nonumber \\
& \Big( r^{(t+1)} + \eta \max_{a'} L'(\theta^{L'}|s^{(t+1)}, a') \Big) 
- L(\theta^{L}|s^{(t)}, a^{(t)})
\label{loss}
\end{align}
where the symbols $L' \in \{Q', \mu'\}$ and $L \in \{Q, \mu\}$ are the target and training networks, respectively. The parameters $\theta^{L}$ and $\theta^{L'}$ represent the related weights, while $r^{(t+1)}$ indicates the reward value calculated at time step $t+1$. The factor $\eta$ represents the reward discount coefficient. 

\begin{algorithm}[!t]
\label{Algo_2}
\DontPrintSemicolon
\small
\SetAlgoCaptionLayout{centerline}
{\caption{\small DDPG-Based Proposed Algorithm}}

\textbf{Initialization:} 
Initialize the critic network $Q$ and the actor network $\mu$ with random parameters $\theta^Q$ and $\theta^\mu$;

\textbf{Initialization:} 
Initialize the target networks $Q'$ and $\mu'$ such that $\theta^{Q'} \gets \theta^Q$ and $\theta^{\mu'} \gets \theta^\mu$;

\textbf{Initialization:} 
Create an empty replay buffer $F$;

\For{episode $= 1$ to $M$}{
    Initialize the initial state $s_1$;\;
    
    \For{$t = 1$ to $T$}{
        Select an action $a^{(t)}$ using the actor network $\mu$, followed by normalization of the beamforming vectors;\;
        
        Execute action $a^{(t)}$ and observe the reward $r^{(t+1)}$ along with the next state $s^{(t+1)}$;\;
        
        Store the transition $(s^{(t)}, a^{(t)}, r^{(t+1)}, s^{(t+1)})$ in the replay buffer $F$;\;
        
        Compute the Q-value using the critic network;\;
        
        Randomly sample a minibatch of experiences from $F$;\;
        
        Update the critic network $Q$ by minimizing the loss function in \eqref{loss};\;
        
        Update the actor network $\mu$ using the sampled policy gradient;\;
        
        Soft-update the target networks as follows: 
        $\theta^{Q'} \gets \tau \theta^Q + (1 - \tau)\theta^{Q'}$, 
        $\theta^{\mu'} \gets \tau \theta^\mu + (1 - \tau)\theta^{\mu'}$;\;
    }
}
\end{algorithm}

\subsection{DDPG Framework}

The DDPG environment can be described as follows:
\begin{itemize}
  \item The action at time step $t$ is defined as
  \begin{align}
  a^{(t)} = \left[\boldsymbol{w}_k^{(t)}, p_{1,1}^{(t)}, \ldots, p_{n_b,p}^{(t)}, \ldots, p_{N,P_n}^{(t)} \right],
  \end{align}
  which includes the beamforming vector as well as the pinching locations along the waveguides.

  \item The state representation is given by
  \begin{align}
  s^{(t)} = \Big[ &a^{(t-1)}, \|\boldsymbol{w}_k^{(t-1)}\|^2,      (\textbf{g}_1^{(t)})^T, \dots,(\textbf{g}_{N_b}^{(t)})^T, (\textbf{h}_{1,k}^{(t)})^T,\nonumber \\
  &\dots,(\textbf{h}_{N_b,k}^{(t)})^T \Big],
  \end{align}
  which encompasses the previous action, the power of beamforming vector dedicated to the user $k$, in-waveguide propagation channel, and the wireless channel from the waveguides to the user $k$.

  \item The reward function is formulated in the following manner:
  \begin{align}
  &r^{(t)} = R_k^{(t)} 
  +  \nonumber \\
  & I_1\left( R_k^{(t)} - R_{th} \right)\cdot pen1 - \sum_{p=1} ^{P_n} \sum_{n = 1}^N I_2(p_{n,p}^{(t)})\cdot pen2 \nonumber \\
  & - \sum_{n_b = 1}^N \sum_{p=1}^{P_n} \sum_{p' \neq p} ^{P_n} I_3\left( p_{n,p}^{(t)} - p_{n,p'}^{(t)}\right)\cdot pen3,
  \end{align}
  with the goal of maximizing the rate while guaranteeing the QoS requirement for the movable user and ensuring that the pinching locations are within a predefined area along the waveguide with a minimum spacing of $D$ between any two pinching spots. 
\end{itemize}
The indicator functions $I_1(.)$, $I_2(.)$, and $ I_3(.)$ can be defined as follow:
    \begin{equation}
I_1(x)  = \begin{cases}0, & \mbox{if } \mbox{x $\geq$ 0}, \\  x, & \mbox{if } \mbox{x $<$ 0}, \end{cases}
\end{equation}

    \begin{equation}
I_2(x)  = \begin{cases}0, & \mbox{if } \mbox{x $\in$ $\mathcal{D}_{n}$}, \\  x, & \mbox{otherwise}, \end{cases}
\end{equation}

    \begin{equation}
I_3(x)  = \begin{cases}0, & \mbox{if } \mbox{x $\geq$ $D$}, \\  x, & \mbox{if } \mbox{x $<$ $D$}. \end{cases}
\end{equation}
These indicator functions are designed in order to impose the penalty terms $pen1$, $pen2$, and $pen3$ when the constraints \eqref{c1} - \eqref{c3} are violated. $I_1$ will be activated if the QoS requirement of the user $k$ is not met. On the other hand, $I_2$ and $I_3$ will be activated if the pinching position is selected outside the waveguide area or the minimal distance $D$ between any two pinching locations is not respected. Now, that  handling the constraints \eqref{c1} - \eqref{c3} is done, the power budget constraint \eqref{c0} is tackled by applying a normalization strategy \cite{9110869}. Let $\boldsymbol{w}_k^{(n)}$  denote the beamforming vector of user $k$ at the $n$-th iteration of the training algorithm at the BS. Accordingly, the beamforming vector $\boldsymbol{w}_k^{(n)}$ is updated as follows:
\begin{align}
&\boldsymbol{w}_k^{*(n)} = \boldsymbol{w}_k^{(n)}\sqrt{\frac{P_{BS}}{P_t^{(n)}}},
\label{normall}
\end{align}
where $P_t^{(n)}$ can be expressed as:
\begin{align}
&P_t^{(n)} = || w_k^{(n)}||^2.
\end{align}

\section{DDPG Algorithm}
The adopted DDPG algorithm is presented in \textbf{Algorithm}~1. First, four networks that encompass the actor--critic architecture are initialized with their respective parameters, together with a replay buffer $F$. At each episode, we reset the environment by initializing new channel conditions, beamforming vector, and location for the movable user. For each $t$ within the total number of episodes, the actor network generates an action, which is then modified by applying the normalization technique described in \eqref{normall} to the beamforming vector. The selected action is applied to the environment and the resulting tuple $(s^{(t)}, a^{(t)}, r^{(t+1)}, s^{(t+1)})$ is added to the replay buffer $F$ to be stored. Next, a random batch of the experiences is then sampled from the replay buffer to update the parameters of the different networks. After completion of the training procedure, the learned model is saved and will be utilized to generate the results during the testing phase.

\section{Computational Complexity Analysis}

\par We denote by $L_1$ and $L_2$ the number of neurons in the first and second hidden layers of the network, respectively. Let $I_{\mu}$ and $U_{\mu}$ represent the number of neurons in the input and output layers of the actor network $\mu$. For the critic network $Q$, the input dimension is given by $I_{Q} = I_{\mu} + U_{\mu}$, while its output layer consists of a single neuron. Consequently, the complexity of the training phase can be denoted as
\[
O\left(MB \left( L_1 (I_{\mu} + I_{Q}) + 2L_1L_2 + L_2 (U_{\mu} + 1) \right) \right),
\]
where $M$ is the total number of training iterations \cite{10636958} and $B$ being the batch size. 
After deployment, the time complexity is expressed as $O(W L N_e)$, where $L$ represents the number of layers in the actor network, $W$ is the number of time steps per episode, and $N_e$ denotes the number of neurons per layer. In contrast, the complexity of the SCA-based approach can be expressed as
\[
O\left(N_{\text{itr}} \cdot J_1 \left( J_2 \log\left(\frac{1}{\epsilon^3}\right) Z^{4.5} \right) \right),
\]
where $N_{\text{itr}}$ is the number of iterations required for exhaustive search, $J_1$ and $J_2$ denote the maximum number of iterations needed for convergence to a predefined threshold, and $\epsilon$ represents the solution accuracy \cite{10109654}. Therefore, comparing the runtime of both methods reveals that the DDPG approach achieves lower computational complexity after deployment, as it leverages knowledge acquired during training. 
\begin{figure}[!t]
    \centering    \includegraphics[width=1\columnwidth]{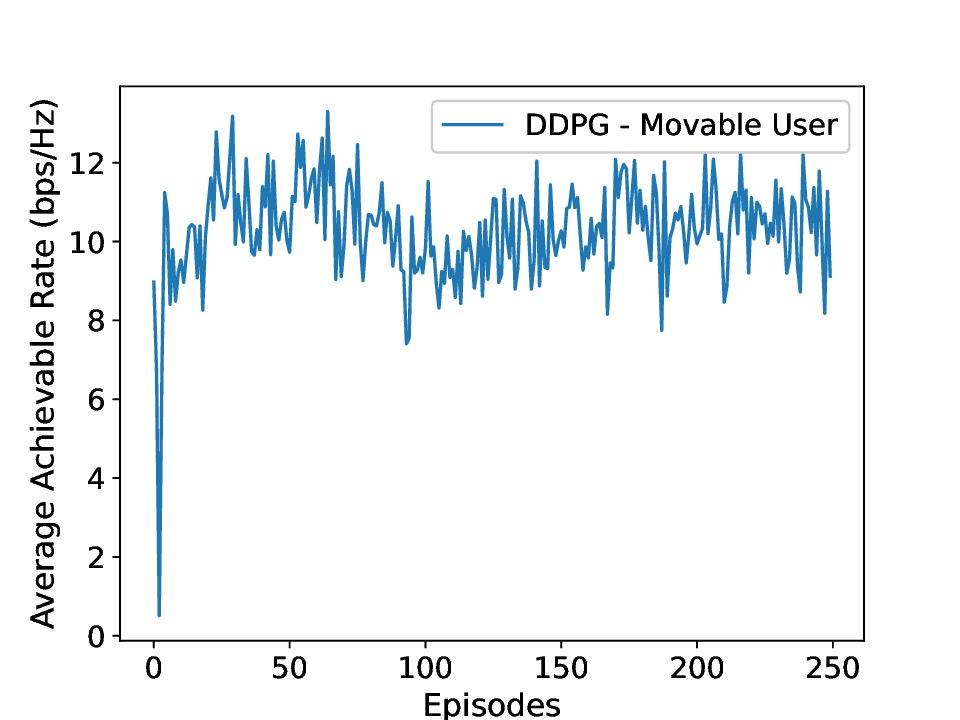}
    \caption{Convergence of the solution approach.}.
	\label{conv}
\end{figure}
\begin{figure}[!t]
    \centering    \includegraphics[width=1\columnwidth]{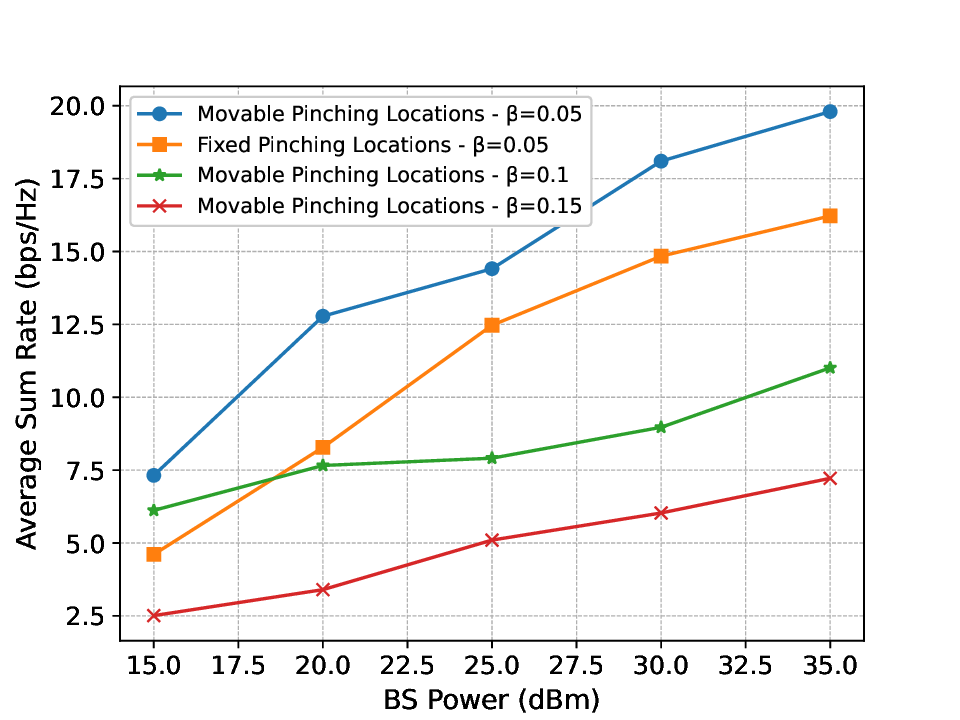}
    \caption{BS power vs average rate.}.
	\label{pbs}
\end{figure}
\begin{figure}[!t]
    \centering    \includegraphics[width=1\columnwidth]{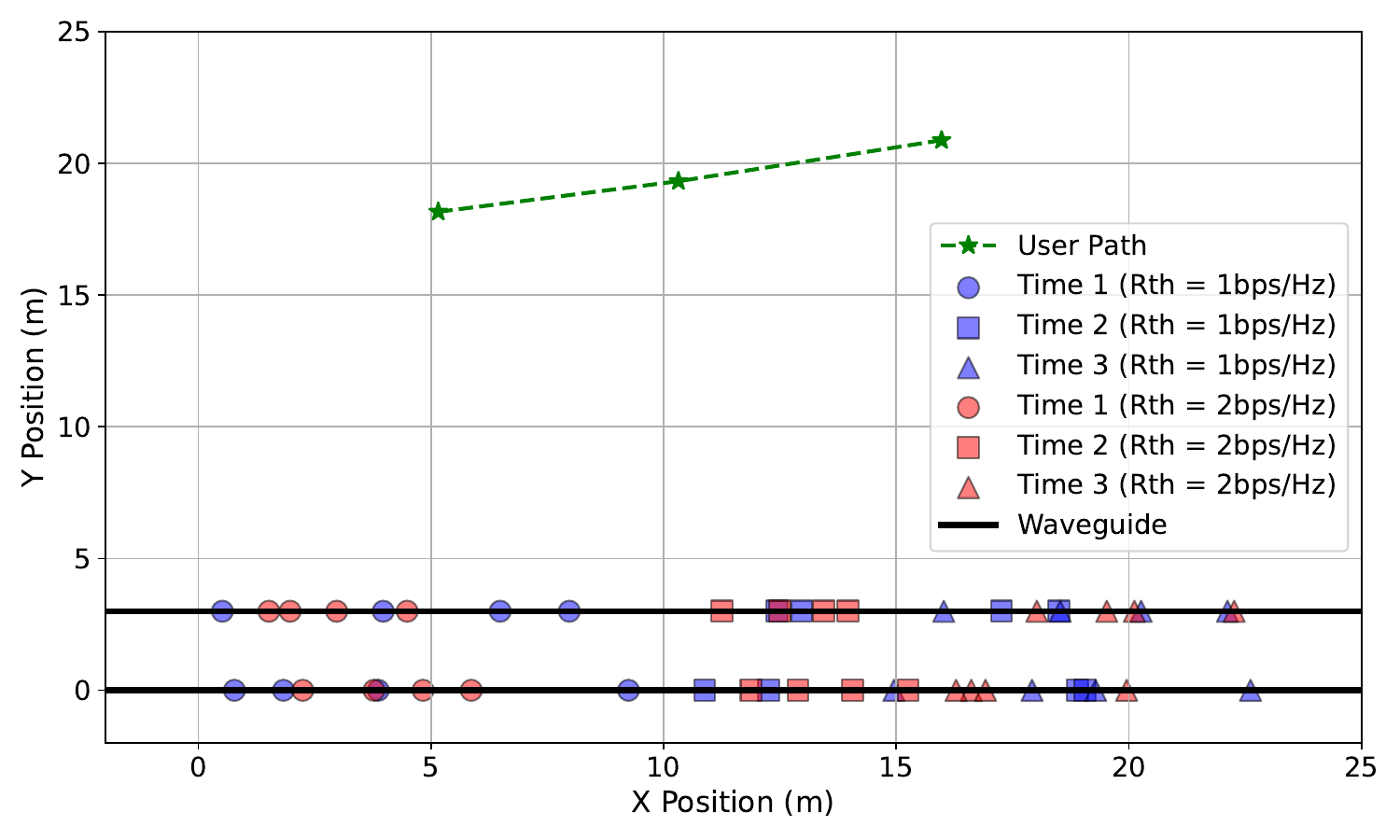}
    \caption{User movement and pinching locations map.}.
	\label{rth}
\end{figure}
\section{Numerical analysis}
\begin{table}[t]
\caption{Simulation Parameters}
\centering
\begin{tabular}{|l|c|l|c|}
    \hline
        Parameters & Values & Parameters & Values \\ \hline
        ~$\beta$ & ~0.05 & ~$N$ & ~2 \\ \hline
        ~$P_{BS}$ & ~20dBm & ~$\sigma^2$ & -174 dBm/Hz \\ \hline
        ~$\mathcal{D}_{1}, \mathcal{D}_{2}$ & ~100 & ~$A_{1}, A_{2}$ & ~10m \\ \hline
         ~$y_{1}, y_{2}$ & ~0, 3m & ~$P_{1}, P_{2}$ & ~4 \\ \hline
        ~$R_{th}$ & ~1bps/Hz & ~$\alpha$ & ~3.9\\ \hline
        ~$\tau_Q$ & 0.001 & ~$\tau_{Q'}$ & ~0.001\\ \hline
        ~$\tau_\mu$ & ~0.001 & ~$\tau_{\mu'}$ & ~0.001\\ \hline
        ~$\lambda$ & ~11.1mm & ~$D$ & ~$\frac{\lambda}{2}$\\ \hline
  \end{tabular} 
\label{T2}
\end{table}
In order to validate the performance of the model under different environment parameters we analyze in this section the numerical results. The simulation parameters adopted in this work are presented in \textbf{Table} 1. Also,  the results are varied over 150 monte-carlo realizations.

Here, it is worth mentioning that we assume that the reconfiguration time of the pinching antennas is negligible compared to the 1 s control interval. This assumption is justified by a separation of timescales: user mobility evolves relatively slowly (maximum speed 5 m/s), while the variation in optimal pinching locations between consecutive time steps remains limited, leading to small, required displacements.
\par Fig.~\ref {conv} demonstrates the convergence analysis of the reinforcement learning framework, where the average reward at the end of each episode is studied versus the episode number. During the first few episodes, the average reward increases rapidly, signifying the ability of the algorithm to learn to enhance the rate by determining the beamforming vector and pinching locations. With the training progression, the reward  eventually stabilizes, demonstrating convergence. This behavior can be explained by the agent’s ability to learn from its interactions with the environment and to generate actions that increasingly satisfy the imposed constraints. 

\par In Fig.~\ref{pbs}, the graphs studies the power budget of the BS vs the achievable sum rate of the user $k$. The performance of the presented model is compared against a benchmark where the pinching locations are distributed in predefined fixed locations along the waveguide with different blockage probabilities. It is clear that with an increase in the invested power budget, the rate showed a significant increase for both models. However, the adopted model proved to be superior reflecting the importance of optimizing the pinching locations as a way to adapt to the movability of the user along with the blockage model. In addition, we can observe that the blockage probability acquires significant effect on the proposed model leading to a decreased average sum rate when the probability increase from $\beta = 0.05$ to $\beta = 0.1$.

\par Fig.~\ref{rth} illustrates the positions of the pinching antennas along the waveguides as a function of the user’s movement under different rate thresholds. It can be observed that, for both QoS requirements, the pinching antennas tend to move closer to the user as it travels from left to right in order to maximize the rate. Moreover, the benchmark with $R_{th}=2$ exhibits a tighter and more concentrated deployment around the user, aiming to further enhance the achievable data rate to satisfy the constraints.
\section{Conclusion}
Within this framework, we considered a scenario where a PASS-enabled BS serves a mobile user and formulated an optimization problem to maximize the average sum rate over a predefined time horizon while satisfying QoS constraints. The problem was shown to be highly non-convex due to the strong coupling among variables, further complicated by user mobility and probabilistic blockage effects. To overcome these challenges, we employed a DDPG approach within a reinforcement learning framework suited for continuous state and action spaces. Simulation results confirmed the effectiveness of the proposed method and how crucial it is to allow real-time adaptation for the pinching locations to fully exploit the potential of PASS.

\bibliographystyle{IEEEtran}
\bibliography{ref}
\end{document}